\documentclass[aps,prd,twocolumn,superscriptaddress,preprintnumbers]{revtex4}

\usepackage[latin1]{inputenc}
\usepackage[dvips]{graphicx,color}
\usepackage{amsmath}
\usepackage{amsfonts}
\usepackage{amssymb}
\usepackage{rotating}
\usepackage{subfigure}

\begin{document}

\preprint{SLAC-PUB-13548}

\title{Bounds on Cross-sections and Lifetimes for Dark Matter Annihilation and Decay into Charged Leptons from Gamma-ray Observations of Dwarf Galaxies}

\author{Rouven Essig}
\affiliation{Stanford Linear Accelerator Center, Stanford, California, 94309, USA}

\author{Neelima Sehgal}
\affiliation{Kavli Institute for Particle Astrophysics and Cosmology, Stanford University, Stanford, California 94305, USA}

\author{Louis E. Strigari}
\affiliation{Kavli Institute for Particle Astrophysics and Cosmology, Stanford University, Stanford, California 94305, USA}

\date{\today}

\begin{abstract}
We provide conservative bounds on the dark matter cross-section and lifetime from
final state radiation produced by annihilation or decay into charged leptons, either directly
or via an intermediate particle $\phi$.
Our analysis utilizes the experimental gamma-ray flux upper limits from four Milky Way dwarf
satellites: HESS observations of Sagittarius and VERITAS observations of Draco, Ursa Minor,
and Willman 1.  Using $90\%$ confidence level lower limits on the integrals over the dark matter distributions,
we find that these constraints are largely unable to rule out dark matter annihilations or decays as an explanation
of the PAMELA and ATIC/PPB-BETS excesses.
However, if there is an additional Sommerfeld enhancement in dwarfs, which have a  velocity dispersion $\sim 10$ to 20 times lower
than that of the local Galactic halo, then the cross-sections
for dark matter annihilating through $\phi$'s required to explain the excesses are very close to the
cross-section upper bounds from Willman 1.
Dark matter annihilation directly into $\tau$'s is also marginally ruled out by Willman 1 as an
explanation of the excesses, and the required cross-section is only a factor of a few below the
upper bound from Draco.
Finally, we make predictions for the gamma-ray flux expected from
the dwarf galaxy Segue 1 for the Fermi Gamma-ray Space Telescope.  
We find that for a sizeable fraction of the parameter space in which dark matter annihilation
into charged leptons explains the PAMELA excess, Fermi has good prospects
for detecting a gamma-ray signal from Segue 1 after one year of observation.
\end{abstract}


\maketitle

\section{Introduction}

While the evidence for non-baryonic dark matter is compelling on astrophysical scales, its microscopic nature remains unknown.  One appealing
possibility is that dark matter consists of particles interacting via the weak force (WIMPs).  Such particles have an annihilation cross-section at freeze-out of $\sim 3 \times 10^{-26}$ cm$^{3}$s$^{-1}$, and if these particles are in the mass range of about $10$ GeV to $10$ TeV, then this would yield the correct relic abundance of dark matter observed today.  In addition, the hierarchy problem of particle physics suggests new physics beyond the Standard Model should enter on scales of 100 GeV to 1 TeV, providing additional motivation for the WIMP hypothesis.  If dark matter consists of WIMPs, then it could annihilate or possibly decay into lighter particles which could be observed as gamma-rays, cosmic rays, or neutrinos.  There are a host of current and planned instruments that have the potential to detect this signal via an observed excess of these annihilation or decay products.  Such indirect search techniques complement direct dark matter detection experiments and attempts at accelerator-driven dark matter production.

Recent results from the PAMELA satellite indicate an excess of positrons between 10 GeV and 80 GeV \cite{Adriani2008},
but no excess in anti-protons \cite{Adriani:2008zq}, above the background expected from cosmic ray nuclei interactions with the interstellar medium.
This confirms earlier hints of a positron excess seen by other experiments (e.g., \cite{Boezio2000,Beatty:2004cy,Aguilar:2007yf,Grimani:2002yz}).
The balloon-borne experiment ATIC recently reported an excess of cosmic ray electrons plus positrons in the energy range of 300-800 GeV \citep{Chang2008}, confirming earlier excesses seen by PPB-BETS \cite{Torii:2008xu}.  Since cosmic ray electrons and positrons loose energy rapidly through inverse Compton and synchrotron processes, the source of these excesses should be less than 1 kpc away \cite{Kobayashi:2003kp}.  It is possible for these excesses to be explained by conventional astrophysical objects such as nearby pulsars \cite{Harding1987,Atoyan:1995ux,Hooper:2008kg,Yuksel:2008rf,Profumo:2008ms}.   It is also possible that they may be the first indirect evidence of dark matter annihilations (e.g., \cite{Cholis2008,Cirelli:2008pk,Arkani-Hamed2008,Fairbairn:2008fb,Nelson:2008hj,Nomura:2008ru,Harnik:2008uu,Bai:2008jt,Bertone2008,Fox:2008kb,Zurek:2008qg,Chun:2008by,Allahverdi:2008jm,Hooper:2008kv,Chen:2008fx,Gogoladze:2009kv,Mardon2009,Meade2009,Barger:2008su,Ibe:2008ye,Ibe:2009dx,Park:2009cs,Huh:2008vj,Bae:2008sp,Allahverdi:2009ae,Feldman:2008xs})
or decays (e.g., \cite{Yin:2008bs,Chen:2008qs,Nardi2008,Ishiwata:2008qy,Ishiwata:2008cv,Arvanitaki2008,Hamaguchi:2008ta,Takahashi:2009mb,Goh:2009wg,Kyae:2009jt,Bae:2009bz}).

In light of the latter possibility, it has been shown that both PAMELA and ATIC/PPB-BETS, as well as a possible excess in microwave photons at the Galactic center dubbed the WMAP haze \cite{Finkbeiner:2003im,Dobler:2007wv,Finkbeiner:2004us,Hooper:2007kb}, may be explained by a $\sim 1$ TeV dark matter particle which annihilates into charged leptons \cite{Cirelli:2008pk,Cholis2008}.  However, for this scenario to work, the annihilation cross-section in the Galactic halo must be boosted by factors of  ${\cal O}(100)$ beyond the freeze-out cross-section.  One suggestion is that this could be achieved via a velocity dependent cross-section through the mechanism of Sommerfeld enhancement \cite{Cirelli:2008pk,Arkani-Hamed2008, Pospelov:2008jd,MarchRussell:2008yu,MarchRussell:2008tu}.  This idea supposes that the ${\cal O}(100)$ lower velocity dispersion in the Galactic halo as compared to the velocity at freeze-out could account for the necessary boost to explain these recent observations.

The main signal of interest in this work is the gamma-ray flux from final state radiation (FSR) by charged leptons produced in dark matter annihilations or decays.  Here FSR is defined as photons directly radiated from the external legs (charged leptons) of the interaction.  This signal has the advantage that the gamma-ray spectrum has a sharp cutoff at energies equal to the dark matter mass and that the normalization and shape of the spectrum can be predicted without knowledge of the astrophysical environment \citep{Birkedal2005}.  In addition, for dark matter masses of a few hundred GeV, the FSR signal is relevant at high photon energies ($> 100$ GeV) making it accessible to Atmospheric Cherenkov Telescopes (ACTs).

We focus on dwarf galaxies because the largest FSR signal is expected from regions with the highest dark matter density.  The Galactic center is a promising source, however it is difficult to separate dark matter produced gamma-rays from other astrophysical backgrounds toward the Galactic center, and there is wide uncertainty in the shape of the central dark matter density profile.  However, dwarf galaxy satellites of the Milky Way are relatively nearby, are highly dark matter dominated with mass to light ratios of ${\cal O}(10-100)$, and are largely free of astrophysical backgrounds as they have little warm or hot gas, minimal dust, and no magnetic fields~\cite{Mateo:1998wg}.  In addition, their dark matter distributions can be inferred directly from stellar kinematics, and their velocity dispersions are of ${\cal O}(10)$ less than in the Galactic halo (see e.g. \cite{Strigari2007}), the latter of which could result in an order of magnitude larger boost of the annihilation cross-section if Sommerfeld enhancement is relevant.

In Section \ref{sec:background}, we discuss the calculation of the gamma-ray signal from dwarf galaxies and provide the formulae we use to
calculate the FSR signal.  Section \ref{sec:ACTs} reviews the most recent gamma-ray constraints from the ACTs HESS and VERITAS.  In Sections \ref{sec:annihilationbounds} and \ref{sec:decaybounds}, we give conservative bounds on the dark matter annihilation cross-sections and decay lifetimes, respectively, for the channels in which dark matter goes to $e^{+}e^{-}$ or $\mu^{+}\mu^{-}$ either directly, or through an intermediate particle (resonance), which we call $\phi$.  We also consider the case in which dark matter goes directly to $\tau^{+}\tau^{-}$.  These bounds are implied by gamma-ray observations of the dwarf galaxies Sagittarius, Willman 1, Draco, and Ursa Minor.  We use updated stellar kinematic data to calculate the dark matter density profiles of these dwarf galaxies, marginalizing over a wide range of profile shapes.  We also discuss the implications of our constraints for the dark matter explanation of the PAMELA and ATIC/PPB-BETS excesses.  In Section \ref{sec:Fermipredictions}, we discuss the prospects for the Fermi gamma-ray space telescope to detect a dark-matter-induced gamma-ray signal from the dwarf galaxy Segue 1.  Among the known dwarf galaxies, Segue 1 could potentially provide one of the largest signals from dark matter annihilations \cite{Geha:2008zr}.
We summarize our results in Section \ref{sec:conclusions}.

\section{Gamma-rays from Annihilating and Decaying Dark Matter}\label{sec:background}

The gamma-ray flux from annihilating dark matter in a dark matter halo is given by

\begin{equation}\label{equ:flux}
\frac{dN_{\gamma}}{dAdt} = \frac{1}{8\pi} \mathcal{L}_{\rm{ann}}(\rho^{2}(\vec{r}),D)
\frac{\langle\sigma v\rangle}{m_{\chi}^2} \int_{E_{th}}^{E_{max}}
\frac{dN_{\gamma}}{dE_{\gamma}} dE_{\gamma}
\end{equation}
where $m_{\chi}$  is the dark matter particle mass, $\langle\sigma v\rangle$ is the annihilation cross-section, $E_{th}$ is the threshold energy of a given gamma-ray instrument, and $E_{max}$ is the maximum energy of the photons.  The integral over $\frac{dN_{\gamma}}{dE_{\gamma}} $ depends only on the particle physics details of the dark matter annihilation.  $\mathcal{L}_{\rm{ann}}$ is given by

\begin{equation}
 \mathcal{L}_{\rm{ann}} = \int_{0}^{\Delta\Omega}
\left\{  \int_{LOS} \rho^{2}(\vec{r}) ds \right\} d\Omega
\label{eq:LOS}
\end{equation}
and depends solely on the properties of the dark matter halo and the solid angle over which it is observed.
Here $\rho(\vec{r})$ is the halo dark matter density profile, $r = \sqrt{D^2+s^2-2sD\rm{cos}\theta}$ for a halo at a distance $D$, and the integral is performed along the line of sight over a solid angle $\Delta \Omega = 2\pi (1-\cos \theta)$.

We will also be considering constraints on decaying dark matter from FSR. In this case, the $\gamma$-ray flux
is analogous to the annihilating case with the following substitutions:
$\langle\sigma v\rangle/(2M_{\chi}^2)  \to \Gamma/M_{\chi}$ and in $\mathcal{L}$, $\rho^{2} \to \rho$.
We define the equivalent integral to eq.~(\ref{eq:LOS}) over the density for the case of
decays as ${\cal L}_{\rm dec}$.
Since decaying dark matter is sensitive to $\rho$ and not $\rho^2$, the dark matter constraints from gamma-ray observations will be weaker for the case of decays.

\subsection{Dwarf Galaxy Dark Matter Density Integrals}

We determine the line-of-sight integral in eq.~(\ref{eq:LOS}) from the kinematic data in the dwarf galaxies,
in a manner similar to the analysis of~\cite{Strigari:2007at}.  For each galaxy we obtain the distribution
of the factors ${\cal L}_{\rm ann}$ and  ${\cal L}_{\rm dec}$ assuming a five-parameter density profile
of the form
\begin{equation}
\rho(r) = \frac{\rho_0}{\big(\frac{r}{r_0}\big)^a \big(1+\big(\frac{r}{r_0}\big)^b\big)^{\frac{c-a}{b}}}.
\end{equation}
We marginalize over $a,b,c,\rho_0,r_0$, with uniform priors, over conservative and
plausible values for these parameters.  The specific ranges for the shape parameters
are $a = [0-1.5]$, $b = [0.5-1.5]$, $c=[2-5]$.
Though numerical simulations produce cuspy profiles for cold dark matter (CDM) models, 
which profiles have $a\simeq 1$~\cite{Navarro:2008kc}, we allow for a broader range
to account for the fact that the dark matter we consider here may be produced
non-thermally, may be self-interacting, or may be warmer than standard CDM.  In this way, our analysis differs from~\cite{Strigari:2007at}, in which CDM priors and cuspy profiles were assumed.
For Sagittarius, Ursa Minor, and Draco,
the scale radius is fixed to be within the range $r_0 = [0.1-10]$ kpc, while for Willman 1 the
scale radius is fixed to be within the range $r_0 = [0.01-1]$ kpc. For each galaxy, $\rho_0$ is
set by the condition that the mass within 300 pc is fixed to the respective observed
values~\cite{Strigari:2008ib}. It is important to note that even though we assume a wide range for the parameters
that determine the density profile, the velocity dispersion data fix the distribution
of the line-of-sight integral to lie within a relatively narrow range.  This is true
despite the fact that the parameters $(a,b,c,\rho_0,r_0)$ are not 
well constrained in and of themselves. This fact precludes us from showing the
best-fit values for $(a,b,c,\rho_0,r_0)$. Regarding the difference between cored and cupsed models, 
we find that allowing for cored models
only reduces the $90\%$ confidence level lower limits, which we use below, by $\sim 10\%$ for each 
galaxy relative to $a\simeq 1$ models.
This traces back to the fact that the stellar kinematics fixes the normalization of the density and
density-squared for the dwarf dark matter halos, so that varying the inner slope has little impact
on the final results~\cite{Strigari:2007at}. 

The 1-dimensional stellar velocity dispersions, averaged over each of the respective galaxies, are listed in Table~\ref{tab:dwarfs}.  We find that the velocity dispersion of the stars is similar to that of the dark matter for these dwarf galaxies.  Assuming the dark matter velocity dispersions are isotropic, we estimate the 3-dimensional dark matter velocity dispersions to be $\sqrt{3}$ times these values.  

Two galaxies listed in Table~\ref{tab:dwarfs} are worth commenting on. First, the satellite Willman 1 was discovered
in 2004, and while there is evidence that this object is dominated by dark matter, better kinematic measurements are required to provide constraints on its dark matter distribution~\cite{Martin2007}. Second, it is clear that Sagittarius is
currently being tidally-disrupted, and for an equilibrium model to be self-consistent
the stars used in the analysis must be bound to the central core of the galaxy. To conservatively account for this issue, in our
analysis we only use the stars from Ref.~\cite{Ibata:1996dv} within the bound region of $\sim 0.4$ kpc~\cite{Majewski:2003ux}.
The remaining two dwarfs, Ursa Minor and Draco, are conclusively known to be dark matter dominated~\cite{Strigari:2007ma,Walker2007}.

\begin{table*}[h!t]
\begin{ruledtabular}
\begin{center}
\begin{tabular}{lccccc}
Dwarf &  Vel. disp. & $\mathcal{L}_{\rm ann}$ &  $\mathcal{L}_{\rm dec}$ & ACT & Flux limit \\
& [km s$^{-1}$] & 
$\log_{10}[\mathcal{L}_{\rm{ann}}/(\rm{GeV}^2 \rm{cm}^{-5})]$  
&  $\log_{10}[\mathcal{L}_{\rm{dec}}/(\rm{GeV} \rm{cm}^{-2})]$  
& 
& $[\rm{cm}^{-2}\rm{s}^{-1}$] \\
\hline
\hline
Sagittarius & 11.4 & $19.35 \pm 1.66$ &$18.73 \pm 1.44$& HESS & $\Phi(E>250 \; \rm{GeV}) < 3.6 \times 10^{-12}$  \\
Draco & 10.0&$18.63\pm0.60$ & $17.51\pm0.12$ & VERITAS &$\Phi(E>200 \; \rm{GeV}) < 2.4 \times 10^{-12}$\\
Ursa Minor & 9.3& $18.79\pm1.26$ & $17.55\pm0.36$ & VERITAS &$\Phi(E>200 \; \rm{GeV}) < 2.4 \times 10^{-12}$\\
Willman 1& 4.3 & $19.55 \pm 0.98$ & $17.51 \pm 0.84$& VERITAS &$\Phi(E>200 \; \rm{GeV}) < 2.4 \times 10^{-12}$
\end{tabular}
\caption{Line-of-sight stellar velocity dispersions and integrals over the mass density and density squared for four Milky Way dwarf galaxies.  Also given are the gamma-ray flux upper bounds for each dwarf from observations with the two Atmospheric Cherenkov Telescopes HESS \cite{Aharonian2008} and VERITAS \cite{Hui2008}.  $\mathcal{L}$ values given represent on minus off regions as defined in Section \ref{sec:ACTs}, and the error bars represent 90\% confidence level upper and lower bounds.}
\label{tab:dwarfs}
\end{center}
\end{ruledtabular}
\end{table*}

\subsection{Final State Radiation Calculations}

For FSR, for the dark matter annihilation channel $\chi\chi \to f\bar{f}$, where $f$ is a fermion, the photon energy spectrum is given by 
\begin{equation}
\frac{dN_{\gamma}}{dy}=\frac{\alpha}{\pi}\left(\frac{1 + (1 - y)^2}{y}\right)\left(\ln\bigg(\frac{s (1 - y)}{m_{f}^2}\bigg) - 1 \right)
\label{equ:fsr1}
\end{equation}
\citep{Beacom:2004pe,Birkedal2005,Mack:2008wu,Bell2008,Cholis2008}.
Here $\alpha\simeq 1/137$, $y=E_{\gamma}/m_{\chi}$, $s=4 m_{\chi}^2$, and $E_{max}=m_\chi$ in eq.~(\ref{equ:flux}).
This formula holds in the collinear limit, where the photon is emitted collinearly with one
of the leptons and $m_f\ll m_\chi$.
For the decay channel $\chi \to f\bar{f}$, we have instead $y=2 E_{\gamma}/m_{\chi}$, $s=m_{\chi}^2$, and $E_{max}=m_\chi/2$.

To calculate the FSR for the channel $\chi\chi \to \phi\phi \to f\bar{f}f\bar{f}$ \cite{Birkedal2005,Cholis2008,Mardon2009},
where $\phi$ is a light particle (resonance),
we start with eq.~(\ref{equ:fsr1}), where now $s=m_\phi^2$ and $y$ is replaced by $x=2E_{\gamma}/m_\phi$.  This gives
the FSR energy spectrum in the rest frame of each $\phi$ for the decay $\phi\to f\bar{f}$.
Next, we boost to the rest frame of $\chi$, which essentially coincides with the Earth's rest frame.
The FSR photon spectrum in the $\chi$ rest frame is given by
\begin{equation}\label{equ:fsr2a}
\frac{dN_{\gamma}}{dy} = 2 \int_y^{\frac{m_\phi-2m_f}{m_\phi-m_f}} dx \,\frac{1}{x}\, \frac{dN_{\gamma}}{dx},
\end{equation}
which may explicitly be written as
\begin{widetext}
\begin{eqnarray}
\frac{dN_{\gamma}}{dy}&=&\!\frac{2\alpha}{\pi y}
\Bigg[
 y^2 + 2y\Big({\rm Li}_{2}\Big[\frac{m_\phi - 2m_f}{m_\phi-m_f}\Big]-{\rm Li}_{2}[y]\Big)
 + (2-y^2)\ln(1-y) + \Big(\ln\Big[\frac{m_{\phi}^2}{m_{f}^2}\Big]-1\Big)
 \bigg\{ 2 - y^2 + 2y\ln\Big[\frac{(m_\phi-m_f)y}{m_\phi-2m_f}\Big] \nonumber \\
&& \;\;\;\;\;\;\;\;
- \frac{(m_\phi^2-2m_f^2)y}{(m_\phi-m_f)(m_\phi-2m_f)}
 \bigg\}
 - \frac{y}{2m_f^2-3m_\phi m_f + m_\phi^2} \bigg\{
  2m_f^2 \Big(2-\ln\Big[\frac{m_f^2 y^2}{(m_\phi-2m_f)^2(1-y)}\Big]
         \Big) \nonumber \\
&&
\;\;\;\;\;\;\;\;
- 3m_f m_\phi
        \Big(\frac{4}{3}-\ln\Big[\frac{m_f(m_\phi-m_f) y^2}{(m_\phi-2m_f)^2(1-y)}\Big]
        \Big)
  + m_\phi^2
        \Big(1-\ln\Big[\frac{(m_\phi-m_f)^2 y^2}{(m_\phi-2m_f)^2(1-y)}\Big]
        \Big)
  \bigg\}
\Bigg].
\label{equ:fsr2}
\end{eqnarray}
\end{widetext}
Here $y=\frac{E_{\gamma}}{m_{\chi}}$, $m_\phi>2m_f$, and the upper limit on the integral in eq.~(\ref{equ:fsr2a})
corresponds to the maximum photon energy in the collinear limit, which approaches unity for $m_f \ll m_\phi$.
For the decay channel $\chi \to \phi\phi \to f\bar{f}f\bar{f}$, we have instead $y=\frac{2 E_{\gamma}}{m_{\chi}}$.
In eq.~(\ref{equ:flux}), $E_{max}=(\frac{m_{\phi}-2m_f}{m_\phi-m_f})m_\chi$ for annihilating dark matter, and the
corresponding formula for decaying dark matter has $E_{max}=(\frac{m_{\phi}-2m_f}{m_\phi-m_f})\frac{m_\chi}{2}$.
Note that the factor of two appears in eq.~(\ref{equ:fsr2a}) since there are two $\phi$'s produced for each dark
matter annihilation or decay.

An important observation is that smaller values of $m_\phi$ lead to a softer photon energy spectrum.
We note that the FSR formula in eq.~(\ref{equ:fsr1}) receives $\mathcal{O}\big(\big(\frac{2m_f}{m_\phi}\big)^2\big)$
corrections, which are small in the collinear limit but which become $\mathcal{O}(1)$ when $m_\phi$ is of the
same order as $2 m_f$.
We will ignore these corrections, since the only case considered in this paper in which they would be important
is when the dark matter decays or annihilates first into a 250 MeV $\phi$, which in turn decays into muons.
In this case, however, the most important contribution to the photon energy spectrum is not from FSR off the muon, but rather from the radiative decay of the muon, which we discuss below.

Note that we assume there is no contribution from internal bremsstrahlung,
which we differentiate from FSR and define as photons directly radiated from
the mediator(s) of the interaction.  This is because we assume the mediator(s)
of both the annihilation and decay channels are electrically neutral.

For the channels that produce muons, we also include the effects of radiative muon decay, in which we
have the decay channels $\mu^{-}\to e^{-}\nu_{\mu}\bar{\nu}_{e}\gamma$ and $\mu^{+}\to e^{+}\bar{\nu}_{\mu}\nu_{e}\gamma$.
The photon spectrum in the muon rest frame has been calculated in \cite{Kuno:1999jp} in the limit
$r\equiv \frac{m_e^2}{m_{\mu}^2}\ll 1$.
For unpolarized muons, it is given by (see also \cite{Mardon2009})
\begin{eqnarray}\label{equ:muonrestframe}
\frac{dN_{\gamma}}{dw} & = & \frac{\alpha}{3\pi}\frac{1-w}{w}\Big((3-2w+4w^2-2w^3)\ln\frac{1}{r} \nonumber \\
& & -\frac{17}{2}+\frac{23 w}{6} -\frac{101 w^2}{12}+\frac{55 w^3}{12} \nonumber \\
&& +(3-2w+4w^2-2w^3)\ln(1-w)\Big),
\label{equ:fsr3}
\end{eqnarray}
where $w=\frac{2 E_{\gamma}}{m_\mu}$.
For the annihilation or decay channel $\chi (\chi) \to \mu^+\mu^-$, the energy spectrum from radiative muon decay in the $\chi$ rest frame
is to a good approximation obtained from eq.~(\ref{equ:muonrestframe}) as
\begin{equation}
\frac{dN_{\gamma}}{dy} = 2 \int_y^1 dw \frac{1}{w}\frac{dN_{\gamma}}{dw},
\label{equ:fsr4}
\end{equation}
where $y = \frac{E_\gamma}{m_\chi}$ in the case of annihilations, and $y = \frac{2 E_\gamma}{m_\chi}$ in the case of decays.
The factor of two appears in eq.~(\ref{equ:fsr4}) since two muons are produced in each annihilation or decay.

For the annihilation or decay channel via an intermediate step involving $\phi$, i.e.~$\chi (\chi) \to \phi\phi \to \mu^+\mu^-\mu^+\mu^-$,
two boosts are required.  The first boost takes us from the rest frame of the muon to the rest frame of the $\phi$:
\begin{equation}
\frac{dN_{\gamma}}{dx} = 2 \int_{\frac{2x}{1+\beta}}^{{\rm min}(1,\frac{2x}{1-\beta})} \,dw \,\frac{1}{w}\frac{dN_{\gamma}}{dw},
\end{equation}
where $x=\frac{2E_\gamma}{m_\phi}$, and $\beta=\sqrt{1-\frac{4 m_\mu^2}{m_\phi^2}}$ is not necessarily close to one, and
a second boost takes us to the rest frame of the $\chi$:
\begin{equation}
\frac{dN_{\gamma}}{dy} = 2 \int_y^1\, dx \,\frac{1}{x}\frac{dN_{\gamma}}{dx},
\end{equation}
where again $y = \frac{2 E_\gamma}{m_\chi}$ in the case of decays, or $y = \frac{E_\gamma}{m_\chi}$ in the case of annihilations.
Note that for the second boost we assumed $m_\phi\ll m_\chi$.

The contribution to the photon spectrum from the radiative muon decay is subdominant to the FSR off the muon
for $m_\chi\gg m_\mu$, if $\chi$ goes directly to muons, and for $m_\phi\gg m_\mu$, if $\chi$ goes to muons via $\phi$'s.
For example, for $\chi\chi\to \mu^+\mu^-$, the number of photons above 200 GeV coming from radiative muon decay is about
(5\%, 10\%, 20\%) compared to the number coming from FSR off the muon for $m_{\chi} = $ (500, $10^3$, $10^4$) GeV, respectively.
(We chose to consider photons above 200 GeV since this is an appropriate threshold energy for VERITAS.)
However, the radiative muon decay dominates over the FSR off the muon if $m_\phi\sim m_\mu$ \cite{Mardon2009,Meade2009}, and
also produces many photons that are much harder than the FSR contribution.
For example, for $m_\phi=250$ MeV, the number of photons above 200 GeV from radiative muon decay is about 35 and 10 times
larger for $m_{\chi} = $ $10^3$ GeV and $10^4$ GeV, respectively, than the number coming from FSR off the muon calculated
with eq.~(\ref{equ:fsr2a}).
For $m_{\chi}$ lighter than about 725 GeV, for the above case, there are no FSR photons off the muon above 200 GeV.  However, there
are photons above 200 GeV from radiative muon decay for even lower values of $m_\chi$.  Thus, the above-mentioned $\mathcal{O}(1)$ correction to the FSR formula in eq.~(\ref{equ:fsr1}), which formally may be
necessary when $m_\phi\sim m_\mu$, is negligible in this case, since the radiative muon decay dominates the signal.

For the channels that produce $\tau$'s, we only include the cases where $\chi$ annihilates or decays directly into $\tau^+\tau^-$.
In addition to the FSR component, $\tau$ decays produce many $\pi^0$'s which each decay into two photons.
We take the parametrization of the total photon spectrum from \cite{Fornengo:2004kj}, which, for $m_{\chi}=1$ TeV, these authors found to be
from simulations
\begin{equation}\label{eqn:taus}
\frac{dN_{\gamma}}{dy} = y^{-1.31} (6.94 y - 4.93 y^2 - 0.51 y^3) e^{-4.53y}.
\end{equation}
This parametrization is valid down to about $y=0.01$ \cite{Fornengo:2004kj}, where $y$ is defined as above.
Although this is the parametrization of the spectrum for $m_\chi=1$ TeV, to a good approximation the
shape of the spectrum as a function of $y$ does not change much over the whole range of values we consider
for $m_\chi$ (a few hundred GeV to 10 TeV).
In particular, the dominant contribution to the photon spectrum coming from the $\pi^0$ decays remains
virtually unchanged, while the sub-dominant contribution from the FSR component changes only logarithmically as a
function of $m_\chi$, see eq.~(\ref{equ:fsr1}).
Note that we also ignore any uncertainties in extracting the photon spectrum from simulations (see e.g.~\cite{Goh:2009wg}).

\section{Gamma-ray Observations from Atmospheric Cherenkov Telescopes}\label{sec:ACTs}

In Table \ref{tab:dwarfs}, we list the current upper bounds on the gamma-ray flux measured for each dwarf galaxy
from various ACTs.  The flux bound for the Sagittarius dwarf is from HESS using 11 hours
of observation \cite{Aharonian2008}.  The bounds for Willman 1, Ursa Minor, and Draco are from recent VERITAS
observations \cite{Hui2008}.  VERITAS observed Willman 1 for 15 hours and Draco and Ursa Minor for 20 hours each, and found gamma-ray flux upper limits for all three dwarfs at the level of $1\%$
of the Crab Nebula \citep{Hui2008}.  We determine the gamma-ray flux of the Crab Nebula
above 200 GeV, which is the
reported energy threshold of VERITAS, by integrating the power-law fit for the Crab Nebula of
$3.2 \times 10^{-11} (E/\rm{TeV})^{-2.49} \rm{cm}^{-2}\rm{s}^{-1}\rm{TeV}^{-1}$ found by \cite{Hillas1998}.
We list $1\%$ of this value in Table \ref{tab:dwarfs} as the flux bounds for Willman 1, Ursa Minor, and Draco.  Similar flux
upper bounds for Willman 1 were found by the MAGIC telescope over 15.5 hours of
observation~\citep{Aliu2008}.  MAGIC also observed Draco over 7.8
hours~\citep{Albert2008}, and STACEE observed Draco for a total of 10.2 hours \citep{Driscoll2008}.  Both obtained flux bounds for Draco about an order of magnitude larger than VERITAS.

\begin{figure}[t!h]
     \begin{center}
     \includegraphics[width=.48\textwidth]{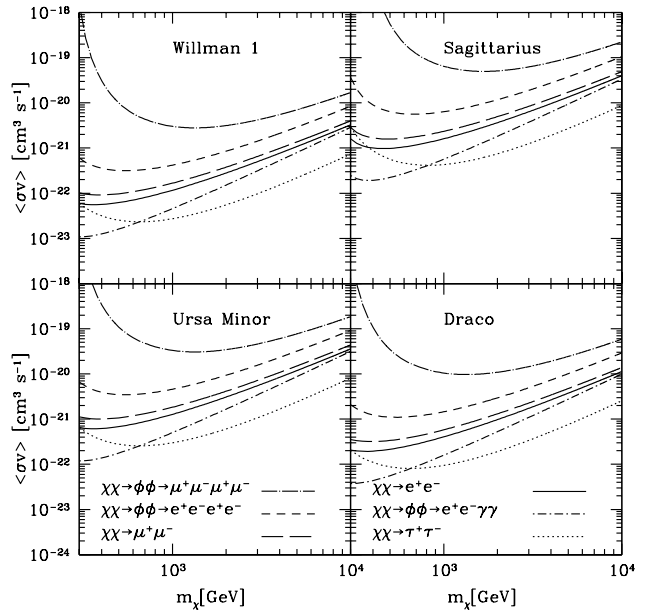}
     \caption{Annihilation cross-section upper bounds as a function of dark matter mass obtained from gamma-ray observations of the four Milky Way dwarf galaxies: Willman 1, Sagittarius, Ursa Minor, and Draco.  These bounds were calculated using the $90\%$ confidence level lower
     limit on each $\mathcal{L}_{\rm{ann}}$ value given in Table \ref{tab:dwarfs}.  The bounds given are for six different channels, which are from top curve to bottom curve: $\chi\chi \to \phi\phi \to \mu^{+}\mu^{-}\mu^{+}\mu^{-}$ for $m_{\phi}=250$ MeV, $\chi\chi \to \phi\phi \to e^{+}e^{-}e^{+}e^{-}$ for $m_{\phi}=100$ MeV, $\chi\chi \to \mu^{+}\mu^{-}$, $\chi\chi \to e^{+}e^{-}$, $\chi\chi \to \phi\phi$ for the case where $\phi$ decays directly to $\gamma\gamma$ with a branching ratio of $10\%$ and to $e^+e^-$ with a branching ratio of $90\%$, and $\chi\chi \to \tau^{+}\tau^{-}$.  Here $\phi$ is a new mediator particle as discussed in e.g.~\cite{Cholis2008,Arkani-Hamed2008}.}
     \label{fig:AnnDarkMatterBounds}
     \end{center}
\end{figure}

To obtain the quantities $\mathcal{L}_{\rm{ann}} $ and $\mathcal{L}_{\rm{dec}}$ for each galaxy,
we must subtract a background level from the signal region, where
both the signal and background regions are specific to each ACT.
In what follows we refer to the background region as the off mode, and the signal region as the on mode.
For HESS, the on region is defined to be within a solid angle of $0.14^\circ$ around the center of Sagittarius,
while the off region is defined as an annulus $0.43-0.57^\circ$ from the center of Sagittarius.
For VERITAS observations, the corresponding on region is within a solid angle of $0.15^\circ$
centered around each dwarf, while the
off region is a ring with inner/outer radius of $0.4-0.5^\circ$.
In principle, one must include the contribution from the diffuse Galactic halo. However, given that we are considering
satellites with angular extent much less than a degree, the contribution from the diffuse Galactic halo emission is similar for both the
on and off regions, so our results do not depend on the normalization of the diffuse halo signal.

Given the on and off regions above, we give the $\mathcal{L}_{\rm{ann}} $ and $\mathcal{L}_{\rm{dec}}$ values for each galaxy in Table~\ref{tab:dwarfs}.
More specifically, these are the respective maximum likelihood ${\cal L}$ values for the on region {\it minus}
the off region. In this way, we determine the distribution
for the best fitting ${\cal L}$ values for each galaxy, fully accounting for correlated error bars.
We note that for the angular regions considered here, the background flux estimated from the off regions equals $\sim 10\%$ and
$50\%$ of the flux within the on regions for annihilation and decays, respectively.
For each entry in Table~\ref{tab:dwarfs}, the error bars represent 90$\%$ confidence levels.

\begin{figure}[t!h]
     \begin{center}
      \includegraphics[width=.48\textwidth]{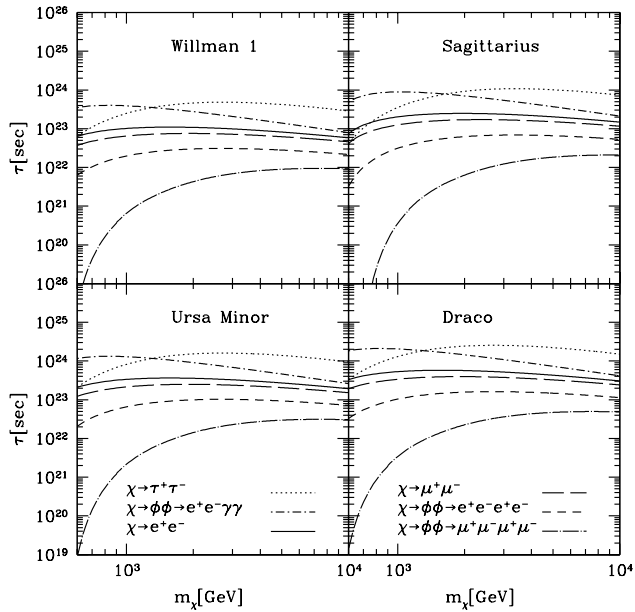}
     \caption{Decay lifetime lower bounds as a function of dark matter mass obtained from gamma-ray observations of the four Milky Way dwarf galaxies: Willman 1, Sagittarius, Ursa Minor, and Draco.  These bounds were calculated using the $90\%$ confidence level lower
     limit on each $\mathcal{L}_{\rm{dec}}$ value given in Table \ref{tab:dwarfs}.  The bounds given are for six different channels, which are from bottom curve to top curve: $\chi \to \phi\phi \to \mu^{+}\mu^{-}\mu^{+}\mu^{-}$ for $m_{\phi}=250$ MeV, $\chi \to \phi\phi \to e^{+}e^{-}e^{+}e^{-}$ for $m_{\phi}=100$ MeV, $\chi \to \mu^{+}\mu^{-}$, $\chi \to e^{+}e^{-}$, $\chi \to \phi\phi \to e^{+}e^{-}$ for the case where $\phi$ decays directly to $\gamma\gamma$ with a branching ratio of $10\%$ and to $e^+e^-$ with a branching ratio of $90\%$, and $\chi \to \tau^{+}\tau^{-}$. }
     \label{fig:DecDarkMatterBounds}
     \end{center}
\end{figure}

\section{Annihilation Cross-section Upper Bounds}\label{sec:annihilationbounds}

Figure \ref{fig:AnnDarkMatterBounds} shows the dark matter annihilation cross-section upper bounds as a function of dark matter mass for
the four dwarf galaxies for six different annihilation channels.  We used the $90\%$ confidence level lower
limit on each  $\mathcal{L}_{\rm{ann}}$ value to give very conservative cross-section upper bounds.
The cross-section upper bounds were calculated using the FSR formulae given in equations (\ref{equ:fsr1}) and (\ref{equ:fsr2}),
including radiative muon decay for channels that include muons.
The weakest bounds are for the $\chi\chi \to \phi\phi \to \mu^{+}\mu^{-}\mu^{+}\mu^{-}$ channel with $m_{\phi} = 250$ MeV.
(The same channel with $m_{\phi} = 1$ GeV (not shown) has a bound that is an order of magnitude stronger at $m_{\chi}=500$ GeV, and
about 25\% stronger at $m_\chi=10$ TeV.)
Stronger constraints in general are found for the channels that annihilate directly into leptons without a $\phi$ mediator, since here the photon spectrum is harder.  However, a strong constraint can be
found for the $\chi\chi \to \phi\phi$ channel for the case where $\phi$ is a scalar of O(100 MeV) and mixes with the Standard Model
Higgs, so that $\phi$ decays to $e^+e^-$ with a branching ratio of about 90\%, and directly to $\gamma\gamma$ with a branching ratio
of about $10\%$ \cite{Gunion:1989we,Cholis2008b,Cholis2008}.
The strongest constraint comes from annihilations into $\tau$'s, due to the large number of high energy photons produced in $\tau$
decays to $\pi^0$'s.
Note that we do not include any boosts from halo substructure here, which may strengthen the cross-section upper
bounds by about a factor of 20 \cite{Martinez2009}.

Among the four dwarfs we consider, the constraints from Willman 1 are the strongest.  Compared to Willman 1, the constraints from Draco, Ursa Minor, and Sagittarius are weaker by a factor of roughly 3.5, 11, and 15, respectively.

We note that in the literature various values of $\mathcal{L}_{\rm{ann}}$ have been used to obtain constraints
for Sagittarius, e.g.~$\mathcal{L}_{\rm{ann}}\simeq$ ($6.4\times 10^{18}$, $4.9\times 10^{19}$, $1.5 \times 10^{21}$) GeV$^2$ cm$^{-5}$
assuming a (large core, Navarro-Frenk-White (NFW) \cite{Navarro:1996gj}, small core) profile, respectively - see e.g.~\cite{Bertone2008,Mardon2009,Meade2009}.
Our central value of $\mathcal{L}_{\rm{ann}}$ for Sagittarius is close to the value quoted for the NFW profile, while the 90\% confidence level lower limit that we use to calculate the upper bounds in Figure \ref{fig:AnnDarkMatterBounds} is about a factor of 13 less than the smallest of the quoted values (large core).

If we assume that the PAMELA, ATIC, and PPB-BETS signals are from dark matter annihilations, then one can calculate the dark matter mass and cross-section for each annihilation channel that best fits these data sets.  To fit the PAMELA data alone, a wide range of $m_{\chi}$ values give a good fit, so we focus on the ATIC/PPB-BETS constraints.  For the channel $\chi\chi \to e^{+}e^{-}$, several authors find a best-fit of $m_{\chi}\sim 700$ GeV and $\langle\sigma v\rangle \sim 6 \times 10^{-24} \ \rm{cm}^{3} \rm{s}^{-1}$ \cite{Cholis2008,Bertone2008,Mardon2009}.  This cross-section is about one order of magnitude smaller than the conservative upper bounds shown in Figure \ref{fig:AnnDarkMatterBounds} for Willman 1.   For the channel $\chi\chi \to \mu^{+}\mu^{-}$, the best-fit value of $\langle\sigma v\rangle \sim 5 \times 10^{-23} \, \rm{cm}^{3} \rm{s}^{-1}$ for $m_{\chi}\sim 1.5$ TeV \cite{Cholis2008,Bertone2008,Mardon2009}, is a factor of $\sim 5$ smaller than the upper bounds from Willman 1.

If we have a $\phi$ mediating the interaction, then the channel $\chi\chi \to \phi\phi \to e^{+}e^{-}e^{+}e^{-}$ gives best-fit values to the ATIC/PPB-BETS data of $m_{\chi}\sim 1$ TeV and $\langle\sigma v\rangle \sim 1 \times 10^{-23} \ \rm{cm}^{3} \rm{s}^{-1}$ \cite{Cholis2008,Mardon2009,Meade2009}.  This cross-section is a factor of $\sim 40$ below the Willman 1 bounds.  For the case where $\phi$ decays to $\gamma\gamma$ with a $10\%$ branching ratio, the best-fit cross-section is only a factor of $\sim 5$ below that from Willman 1.  Note that if Sommerfeld enhancement does not saturate at the local velocity dispersion of the Galactic halo, the lower velocity dispersions found in dwarf galaxies could boost the cross-section by a factor of $v_{\rm{local}}/v_{\rm{dwarf}}\sim$ 10 to 20 depending on the dwarf.  Both Willman 1 and Draco would then disfavor this latter channel as an explanation of the PAMELA and ATIC signals.  For the $\chi\chi \to \phi\phi \to \mu^{+}\mu^{-}\mu^{+}\mu^{-}$ channel, the best-fit value of $\langle\sigma v\rangle \sim 7.5 \times 10^{-23} \, \rm{cm}^{3} \rm{s}^{-1}$ for $m_{\chi}\sim 2.5$ TeV \cite{Cholis2008,Mardon2009}, is a factor of $\sim 50$ below the Willman 1 bound.   Again, if Sommerfeld enhancement is relevant at the velocity dispersions of dwarf galaxies, then the best-fit cross-section could be boosted by a factor of $\sim10$ to 20 in the dwarf, making it close to the current upper bounds.

Note that if the annihilation cross-section varies with velocity through a Sommerfeld enhancement, then a more careful treatment of the expected gamma-ray flux would allow for the cross-section to vary as a function of position within the dwarf galaxy \cite{Robertson:2009bh}.  This would entail including $\langle\sigma v\rangle$ within the integral over the dark matter distribution in eq.~(\ref{eq:LOS}).  We find that using this more careful analysis gives expected boost factors of the same order of magnitude, within the solid angles of interest here, as we get from a more naive treatment using the ratio of the average local and dwarf velocity dispersions.

Although channels which annihilate into $\tau$'s do not give a very good fit to the PAMELA/ATIC/PPB-BETS data \cite{Cholis2008}, it may still be consistent to fit the data with, for example, a 1, 2, or 3 TeV dark matter particle with a cross-section of about $(3, 10, 20) \times 10^{-23} \, \rm{cm}^3 \rm{s}^{-1}$, respectively  \cite{Cholis2008,Bertone2008}.  These best-fit cross-sections are $\sim 1$ to 2 times larger than the upper bounds from Willman 1 (and only $\sim 2$ to 3 times smaller than the bounds from Draco), marginally ruling out dark matter annihilations into $\tau$'s as a possible explanation of the data.

We emphasize that the 90\% upper bound on $\cal{L}_{\rm ann}$ is larger than the 90\% lower bound by about one to three orders of magnitude, depending on the dwarf.  It is thus perfectly conceivable that further ACT observations will see a gamma-ray signal from at least some of these dwarfs if
dark matter is responsible for the PAMELA/ATIC anomalies.

\section{Decay Lifetime Lower Bounds}\label{sec:decaybounds}

In Figure \ref{fig:DecDarkMatterBounds}, we give the lower bounds on the dark matter decay lifetime as a function of dark matter mass.  Again we used the $90\%$ confidence level lower limit for each $\mathcal{L}_{\rm{dec}}$ value to give very conservative bounds.  The channels are the same as those given in Figure \ref{fig:AnnDarkMatterBounds}, except that one dark matter particle decays into leptons or $\phi$'s.  For decays, the 90\% lower
bounds are strongest from Draco, with Ursa Minor and Sagittarius providing the same constraints within a factor of a few.  Willman 1 has about a factor of 5 weaker constraints than Draco.

\begin{figure}[t!h]
     \begin{center}
      \includegraphics[width=.48\textwidth]{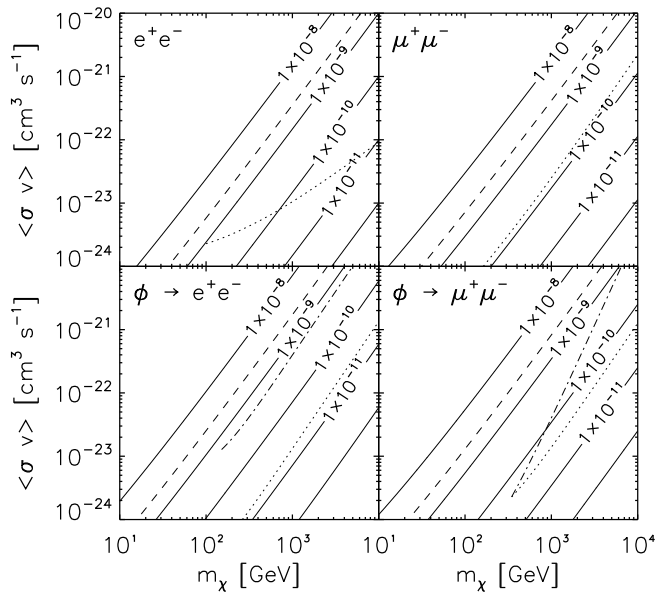}
     \caption{Predicted flux levels (solid black lines) for the Fermi satellite from the dwarf galaxy Segue 1, as a function of dark matter mass and annihilation cross-section, for four different annihilation channels: $\chi\chi \to e^{+}e^{-}$, $\chi\chi \to \mu^{+}\mu^{-}$, $\chi\chi \to \phi\phi \to e^{+}e^{-}e^{+}e^{-}$ for $m_{\phi}=5$ MeV, and $\chi\chi \to \phi\phi \to \mu^{+}\mu^{-}\mu^{+}\mu^{-}$ for $m_{\phi}=250$ MeV.  Here we assume $E_{th}=100$ MeV in eq.~(\ref{equ:flux}), and with this $E_{th}$, after one year of observing, Fermi can detect sources above a flux level of $2.4\times 10^{-9}$ cm$^{-2}$s$^{-1}$ with $3\sigma$ significance (dashed line).  The fluxes were calculated assuming the mean expected value of $\cal{L}_{\rm ann}$.  For all four channels, the dotted lines indicate the approximate values for the annihilation cross-section and mass suggested by the PAMELA data (the ATIC preferred region is, very roughly, a subset of this line).  For the channels with an intermediate $\phi$, the dot-dashed line is the PAMELA suggested region assuming an additional maximum Sommerfeld enhancement to the cross-section in Segue 1.  The expected fluxes for the other dwarf galaxies may be estimated by a simple rescaling with appropriate ratios of $\cal{L}_{\rm ann}$. }
     \label{fig:FermiSegue_100MeV}
     \end{center}
\end{figure}

If the PAMELA, ATIC and PPB-BETS signals are from dark matter decays, then one can calculate the dark matter mass and decay lifetime that best fits the data.  For the channel $\chi\chi \to e^{+}e^{-}$, values of $m_{\chi}\sim 1.5$ TeV and $\tau \sim 10^{26}$ sec are found to best fit the data sets \cite{Nardi2008,Arvanitaki2008}.  This decay lifetime is 2 to 3 orders of magnitude above the lower bounds given by the dwarf galaxies, depending on the dwarf.  For the $\chi\chi \to \mu^{+}\mu^{-}$ channel, best-fit values of $m_{\chi}\sim 2$ TeV and $\tau \sim 10^{26}$ are found \cite{Nardi2008,Arvanitaki2008}, again 2 to 3 orders of magnitude above the lifetime lower bounds given by the dwarfs.  Best-fit values for $\chi\chi \to \tau^{+}\tau^{-}$ are $m_{\chi}\sim 5$ TeV and $\tau \sim 5 \times 10^{25}$ \cite{Nardi2008,Arvanitaki2008}, where the lifetime is about a factor of 20 above the current bounds from Draco.  We see that the gamma-ray constraints from dwarf galaxies do not currently constrain the parameter region of interest for decays.  However, the 90\% upper bound on $\cal{L}_{\rm dec}$ can be an order of magnitude larger than the 90\% lower bound that we used to set these constraints (depending on the dwarf), so that the gamma-rays from decays to, for example, $\tau's$ could conceivably be seen from dwarfs.

\section{Predictions for Fermi}\label{sec:Fermipredictions}

In this section, we discuss the prospects for the Fermi Gamma-ray Space Telescope to detect gamma-ray signals
from dwarf galaxies, assuming that the dark matter annihilates into charged leptons.  Previous work on the Fermi detection
prospects of a gamma-ray signal from dark matter annihilations or decays include e.g.~\cite{Moiseev:1999ft,Baltz:2008wd,Arvanitaki2008},
and e.g.~\cite{Evans:2003sc,Colafrancesco:2006he,Pieri2009} in the context of dwarf galaxies.

\begin{figure}[t!h]
     \begin{center}
      \includegraphics[width=.48\textwidth]{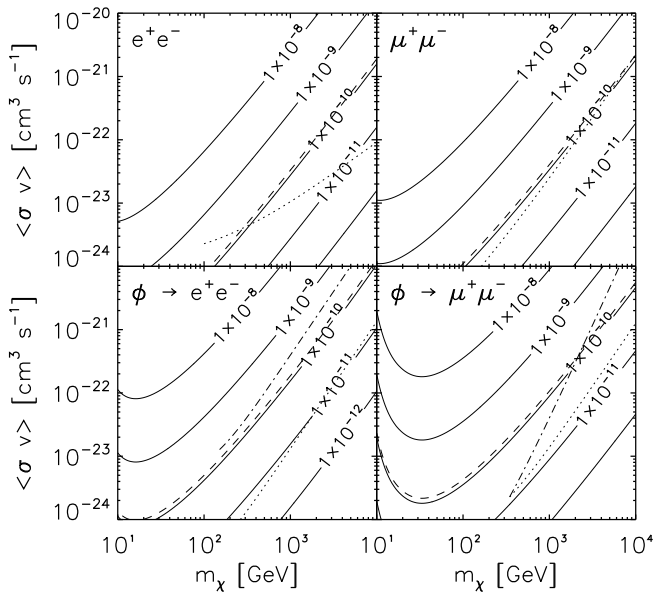}
     \caption{Predicted flux levels (solid black lines) for the Fermi satellite from the dwarf galaxy Segue 1, as a function of dark matter mass and annihilation cross-section, for the four annihilation channels listed in Figure \ref{fig:FermiSegue_100MeV}.   Here we assume $E_{th}=5$ GeV in eq.~(\ref{equ:flux}), and with this $E_{th}$, after one year of observing, Fermi can detect sources above a flux level of $1.2\times 10^{-10}$ cm$^{-2}$s$^{-1}$ with $3\sigma$ significance (dashed line). The fluxes were calculated assuming the mean expected value of $\cal{L}_{\rm ann}$.  The dotted and dot-dashed lines are the same as in Figure \ref{fig:FermiSegue_100MeV}.}
     \label{fig:FermiSegue_5GeV}
     \end{center}
\end{figure}

A disadvantage of the Fermi Large Area Telescope (LAT) over the ACTs is that its effective area
is only $\sim$ 1 m$^2$ \cite{Atwood:2009ez} compared with the effective area of the ACTs which is $\sim 10^5$ m$^2$.
However, an advantage of the Fermi LAT over the ACTs, besides covering the whole sky, is that the LAT will detect photons
with energies down to about 100 MeV \cite{Atwood:2009ez}, while the ACTs only detect photons that are above
around a hundred GeV.
Since the photons expected from dark matter annihilations are never more than the mass of the dark matter,
Fermi would in principle be able to detect lower mass dark matter particles than the ACTs.
Fermi would become especially important compared to the ACTs in the case that only the PAMELA data is caused by dark matter and
the explanation of the ATIC/PPB-BETS data is due to something else.
This is because a dark matter particle with a mass of $\mathcal{O}$(100 GeV) can explain the PAMELA data, whereas a mass above around 700 GeV is required to explain the combined PAMELA/ATIC/PPB-BETS data.

Here, we will present the detection prospects for one recently discovered particularly promising dwarf galaxy,
Segue 1~\cite{Geha:2008zr}.
This satellite is at a distance of 23 kpc from the Sun at a galactic latitude of
$\sim 50.4^\circ$ and has an observed half-light radius of $\sim 30$ pc~\cite{Belokurov:2006ph}.   
It has a measured 1-dimensional stellar velocity dispersion of $\sim 4.3$ km/s ~\cite{Geha:2008zr}.
Segue 1 has only 24 stars with measured line-of-sight velocities~\cite{Geha:2008zr}, so that its ${\cal L}$ value
will be less constrained than any of the previous four objects we have considered. Using the same
formalism as in Section \ref{sec:background}, we find for Segue 1:
\begin{equation}\label{equ:Segue}
{\cal L}_{\rm ann, Seg}  = 10^{20.17 \pm 1.44} \; {\rm GeV^2 cm^{-5}}.
\end{equation}
Here we have assumed an on region of $0.2^\circ$ centered on
Segue 1, and have not performed a background subtraction as
above when considering ACT limits.
This is appropriate when discussing predictions for Fermi as there is no background subtraction in the same fashion as with the ACTs.

In Figure \ref{fig:FermiSegue_100MeV}, we show with solid black lines the expected
gamma-ray flux from Segue 1 as a function of the annihilation cross-section and dark matter mass for the
dark matter annihilation channels $\chi\chi\to e^+e^-$, $\chi\chi\to \mu^+\mu^-$,
$\chi\chi\to \phi\phi\to e^+e^-e^+e^-$ (with $m_\phi=5$ MeV), and $\chi\chi\to \phi\phi\to \mu^+\mu^-\mu^+\mu^-$
(with $m_\phi=250$ MeV).
We took $E_{th}=100$ MeV in eq.~(\ref{equ:flux}), and used the central value of
${\cal L}_{\rm ann, Seg}  = 10^{20.17} \, {\rm GeV^2 cm^{-5}}$
in eq.~(\ref{equ:Segue}).
In Figure \ref{fig:FermiSegue_5GeV}, we show the same channels but with $E_{th}=5$ GeV.
For both figures we assumed that Fermi will be able to detect and reconstruct photons up
to 1 TeV in energy.

To estimate the ability of Fermi to detect gamma-rays from a dwarf galaxy requires a careful analysis of the
response function of the LAT to the particular gamma-ray signal from the dwarf.
Such a careful analysis is beyond the scope of this work, and we content ourselves with a more simple estimate of
Fermi's sensitivity, which is possible with a few reasonable assumptions.
First, we assume that the dwarf is a point source.
Next, we assume that the publicly available point-source sensitivity as a function of galactic latitude
and threshold energy is also applicable in this case, even
though it was calculated assuming a $1/E^2$ spectrum at the source \cite{Atwood:2009ez}
(the FSR signal in eq.~(\ref{equ:fsr1}) has a dependence more like $1/E$).
With these assumptions, the flux sensitivity for Fermi to detect a point source at high galactic latitudes
with a $3\sigma$ significance after 1 year
of observation is about $2.4\times 10^{-9}$ cm$^{-2}$s$^{-1}$ for $E_{th}=100$ MeV, and about
$1.2\times 10^{-10}$ cm$^{-2}$s$^{-1}$ for $E_{th}=5$ GeV.
The flux sensitivity is better at higher threshold energies, at least up to about 5 GeV, above which
the sensitivity remains roughly constant.
The background to calculate these sensitivities was assumed to be uniform with an integrated gamma-ray flux above 100 MeV given by
$1.5\times 10^{-5}$ cm$^{-2}$s$^{-1}$sr$^{-1}$ and with a spectral index of -2.1 \cite{Atwood:2009ez}.
We indicate the respective sensitivities in Figures $\ref{fig:FermiSegue_100MeV}$ and \ref{fig:FermiSegue_5GeV} with dashed lines.

From Figure \ref{fig:FermiSegue_100MeV}, we see that generally the 5 GeV energy threshold is
able to probe more of the cross-section versus mass parameter space than an energy threshold of
100 MeV.  The only exception to this is for very low dark matter mass (10's of GeV) for the
$\chi\chi\to\phi\phi$ channel, with $\phi\to \mu^+\mu^-$ and $m_\phi\sim m_\mu$.
In this case the higher energy threshold removes more of the signal than the background.

An exciting prospect is that Fermi could detect a gamma-ray signal from Segue 1 
if dark matter annihilations are responsible for the PAMELA signal and/or the ATIC/PPB-BETS signal.
In Figures \ref{fig:FermiSegue_100MeV} and \ref{fig:FermiSegue_5GeV}, the 
dotted lines indicate the approximate values for the annihilation cross-section and mass 
suggested by the PAMELA data for the different channels (the ATIC preferred region is, very roughly, a
subset of this line).  
These regions are based on the results found in \cite{Cholis2008}, and include an extrapolation to 
larger masses. 
(By including only a line instead of an extended region, we ignore the uncertainties in fitting to the 
PAMELA data - see e.g.~\cite{Bertone2008}.)
For the channel $\chi\chi\to e^+e^-$, an $\mathcal{O}$(100 GeV) dark matter particle has 
an annihilation cross-section suggested by the PAMELA data that is large enough to give a 
gamma-ray signal detectable with Fermi, assuming a 5 GeV threshold energy.
The cross-sections required to explain the PAMELA data with the channel $\chi\chi\to \mu^+\mu^-$ 
are about a factor of two below the one-year $3\sigma$ detection sensitivity for masses of a few hundred GeV, but lie 
right at the detection sensitivity for masses of several TeV.
With a few years of data, the low mass regions can also be detected.
If dark matter annihilations directly into $\tau$'s (not shown) are responsible for the PAMELA and/or ATIC/PPB-BETS 
excesses, then Fermi can detect the resulting gamma-ray signal with about one year of data 
with higher than $5\sigma$ significance. 

For the annihilation channels via $\phi$'s, the gamma-ray signal tends to be smaller so that longer observations
with Fermi may be required.
However, these channels may receive an additional Sommerfeld enhancement
to the annihilation cross-section in Segue 1 (up to a factor of $\sim 30$), compared to the
local annihilation cross-section required to fit the PAMELA data.
We indicate the PAMELA preferred regions including the additional maximum Sommerfeld enhancement 
with dot-dashed lines in Figures \ref{fig:FermiSegue_100MeV} and \ref{fig:FermiSegue_5GeV}.   
With the additional boost in Segue 1, the preferred cross-sections may well be large enough 
for the gamma-ray signal to be detectable by Fermi.  
A subtlety in calculating the maximum additional Sommerfeld boost in Segue 1 is 
that the enhancement saturates for velocities smaller than about $m_\phi/m_\chi$ \cite{Arkani-Hamed2008}.  Saturation does not have to occur 
for the case $\chi\chi\to\phi\phi\to e^{+}e^{-}e^{+}e^{-}$ as
$m_\phi$ can be as small as a few MeV, allowing for a large additional Sommerfeld 
boost at the dwarf for all $m_\chi$ of interest. 
However, for the case $\chi\chi\to\phi\phi\to \mu^+\mu^-\mu^+\mu^-$, $m_\phi$ must be at least a few 
hundred MeV, which suggests that the 
Sommerfeld enhancement would already saturate locally in our halo for 
dark matter masses of only a few hundred GeV. 
However, for larger dark matter masses, an additional enhancement up to a factor of $\sim 30$, may occur in Segue 1 (see Figures 3 and 4).  

We note that the discussion in the previous two paragraphs assumed that the ${\cal L}$ value for
Segue 1 is the central value given in eq.~(\ref{equ:Segue}).
Since this value has a large uncertainty, the prospects for detection may be either
more pessimistic or more optimistic than presented here.
Moreover, note that it is easy to rescale the expected gamma-ray flux from Segue 1 to estimate the 
flux from other dwarf galaxies.
Ignoring the subtleties of the background subtraction, which are small for annihilations, 
the rescaling factor is given by the ratio of a particular dwarf's $\cal{L}_{\rm ann}$ value in Table \ref{tab:dwarfs} to the $\cal{L}_{\rm ann}$ value of Segue 1 in eq.~(\ref{equ:Segue}).
Since $\cal{L}_{\rm ann}$ is generally smaller for the other dwarf galaxies than for Segue 1, 
the prospects are reduced for Fermi to detect a gamma-ray signal from them.  
However, given the uncertainties in the dwarf's dark matter distribution, it is conceivable 
for Fermi to detect a signal from, for example, Sagittarius and Willman 1.  
Moreover, it is also possible that as of yet undiscovered nearby dwarf satellites with 
very large mass-to-light ratios could first be detected with Fermi through their gamma-ray signal 
from dark matter annihilations.  

In addition to determining the total flux above some energy threshold, Fermi may determine 
the differential photon flux $d\Phi/dE$ (which is
proportional to $dN/dE$) as a function of energy.
Determining $d\Phi/dE$ would be helpful in
determining whether a detection is a result of dark matter annihilation or is
from another source.
For example, the photon spectrum from FSR in the channel $\chi\chi\to f\bar{f}$
plotted as $E^2 dN_\gamma/dE$ is unmistakable, since it has an edge at the mass
of the dark matter particle.
A detection of such an edge would provide compelling evidence that the signal
comes from FSR caused by dark matter annihilating into leptons,
in addition to providing a measurement of the dark matter mass.

Finally, we note that Segue 1 also presents an excellent target for ACTs, which in general 
are more sensitive than Fermi in detecting a gamma-ray signal from dark matter annihilations 
in dwarf galaxies \emph{if} the dark matter mass is above the energy threshold of the ACTs.  
An ACT able to achieve the same kind of sensitivity as VERITAS did with observations 
of Willman 1, Draco, and Ursa Minor, could potentially detect a gamma-ray
signal from Segue 1 if the dark matter annihilates into charged 
leptons.  
Segue 1 may be a particularly good target for the MAGIC instrument, so we
give the background subtracted ${\cal L}$ values for Segue 1 relevant for
MAGIC.  These are
\begin{equation}
{\cal L}_{\rm ann}  = 10^{20.04 \pm 1.40} \, {\rm GeV^2 cm^{-5}}
\end{equation}
for annihilations, and
\begin{equation}
\ {\cal L}_{\rm dec} = 10^{17.80 \pm 0.68} \, {\rm GeV cm^{-2}}
\end{equation}
for decays.
These numbers assume an on-region of $0.1^\circ$, and a background
region of the same area centered $0.8^\circ$ away from the center of
the dwarf, which are the on and off regions applicable to the MAGIC telescope \cite{Aliu2008}.

\section{Conclusions}\label{sec:conclusions}

We provided conservative bounds on the dark matter annihilation cross-section and decay lifetime from
final state radiation produced by dark matter annihilation or decay into charged leptons, and
discussed the implications of our constraints for the dark matter explanation of the PAMELA and
ATIC/PPB-BETS excesses.
We used the experimental gamma-ray flux upper limits for four Milky Way dwarf satellites: HESS
observations of Sagittarius and VERITAS observations of Draco, Ursa Minor, and Willman 1.
We find that for the channels where dark matter annihilates either directly into electrons
or muons, or through an intermediate particle (resonance), which we call $\phi$, the cross-section bounds
from the dwarf galaxies do not currently rule out this form of dark matter as an explanation of
the PAMELA and ATIC/PPB-BETS excesses.
However, if the mechanism of Sommerfeld enhancement is relevant for the factor of $\sim 10$ to 20 lower
velocity dispersions in dwarf galaxies as compared to the local Galactic halo, then the cross-sections
of dark matter annihilating through $\phi$'s required to explain the excesses are very close to the
cross-section upper bounds from Willman 1.
Dark matter annihilation directly into $\tau$'s is also marginally ruled out by Willman 1 as an
explanation of the excesses, and the required cross-section is only a factor of a few below the
upper bound from Draco.
Dark matter decays as an explanation of the observed excesses are currently not constrained by
gamma-ray observations of these dwarf galaxies.

We note that since the uncertainty in the gamma ray flux coming from the uncertainty in the dwarf
dark matter distribution is about 1-3 orders of magnitude, depending on the dwarf, we have provided
rather conservative constraints by using the lower value of the expected fluxes.
However, it is possible that further observations by Atmospheric Cherenkov Telescopes
of various dwarf galaxies could see a gamma-ray signal from dark matter annihilations
\cite{Bringmann:2008kj} (see also \cite{Pieri2009}).

We also make predictions of the gamma-ray flux expected for the Fermi satellite from the dwarf
galaxy Segue 1, which may provide one of the strongest signals from dark matter annihilations
in dwarfs, and is thus also an excellent target for the Atmospheric Cherenkov Telescopes.
We find that in a sizeable fraction of the parameter space in which dark matter annihilation 
into charged leptons explains the PAMELA excess, Fermi has good prospects 
for detecting a gamma-ray signal from Segue 1 after one year of observation.  

\vskip 7mm

{\bf Note added:} While this paper was considered for publication in PRD, the 
measurement of the cosmic ray $e^++e^-$ spectrum from 20 GeV to 1 TeV 
with the Fermi Large Area Telescope appeared \cite{Abdo:2009zk}.  
Their results call into question the sharp peak observed by
ATIC/PPB-BETS, but they nevertheless still observe an anomalous excess 
that might be due to dark matter.  
One can again calculate the dark matter mass and cross-section for each annihilation channel that 
best fits both the PAMELA and Fermi data sets (we use the results in \cite{Meade:2009iu} in the following).  
The channel $\chi\chi \to e^{+}e^{-}$ is not a good fit.  
For the channel $\chi\chi \to \mu^{+}\mu^{-}$, the best-fit value of 
$\langle\sigma v\rangle \sim 3 \times 10^{-23} \, \rm{cm}^{3} \rm{s}^{-1}$ for $m_{\chi}\sim 1.4$ TeV 
is a factor of $\sim 8$ smaller than the upper bounds from Willman 1.
For the channel $\chi\chi \to \tau^{+}\tau^{-}$, the best-fit value of 
$\langle\sigma v\rangle \sim 2 \times 10^{-22} \, \rm{cm}^{3} \rm{s}^{-1}$ for $m_{\chi}\sim 3$ TeV 
is a factor of $\sim 2$ larger than the upper bounds from Willman 1, marginally ruling out dark matter annihilations into $\tau$'s as a possible explanation of the data.
For the channels $\chi\chi \to \phi\phi \to e^{+}e^{-}e^{+}e^{-}$ and 
$\chi\chi \to \phi\phi \to \mu^{+}\mu^{-}\mu^{+}\mu^{-}$, the best-fit values are similar to those for 
ATIC/PPB-BETS and the constraints are thus similar to the values given in Section \ref{sec:annihilationbounds}.  

\begin{acknowledgments}
We thank James Bullock, Peter Graham, Manoj Kaplinghat, Philip Schuster, Natalia Toro, Risa Wechsler, and especially Neal Weiner
for useful discussions. We are grateful to Marla Geha and Beth Willman for access to
unpublished data from Willman 1.
We also thank Stefan Funk for providing important information on the HESS and Fermi
experiments, as well as Matthew Wood and Miguel Sanchez-Conde for providing
important information on the VERITAS, and MAGIC experiments, respectively.
RE is supported by the US DOE under contract number DE-AC02-76SF00515.
NS is supported by the U.S. Department of Energy contract to SLAC no. DE-AC3-76SF00515.
LES acknowledges support for this work from NASA through Hubble Fellowship grant
HF-01225.01 awarded by the Space Telescope Science Institute, which is
operated by the Association of Universities for Research in Astronomy, Inc.,
for NASA, under contract NAS 5-26555.
\end{acknowledgments}

\bibliography{refs}

\end{document}